%% file: main.tex
\documentclass{article}

\usepackage{spconf}
\usepackage{hyperref}       
\usepackage{url}            
\usepackage{booktabs}       
\usepackage{nicefrac}       
\usepackage{microtype}      
\usepackage{amsfonts}       
\usepackage{amsmath}
\usepackage{tabularx}
\usepackage{float}
\usepackage{outlines}
\usepackage{subcaption}
\usepackage{algorithm}
\usepackage{algpseudocode}
\usepackage{chngcntr}
\usepackage{graphicx}
\usepackage{xcolor}
\usepackage{setspace}
\usepackage[medium,compact]{titlesec}
\usepackage{enumitem}
\usepackage{pgfplots}
\pgfplotsset{compat=1.13}

\usepackage[square,sort,comma,numbers]{natbib}

\hypersetup{
    colorlinks,
    linkcolor={red!50!black},
    citecolor={blue!80!black},
    urlcolor={blue!80!black}
}

\renewcommand{\paragraph}[1]{\noindent\textbf{#1}\quad}

\DeclareMathOperator*{\argmin}{arg\,min}

\title{Using External Off-policy Speech-to-Text Mappings in Contextual End-to-End Automated Speech Recognition}
\name{David M. Chan$^{\star \dagger}$ \qquad Shalini Ghosh$^{\dagger}$ \qquad Ariya Rastrow$^{\dagger}$ \qquad Bj{\"o}rn Hoffmeister$^{\dagger}$}
\address{$^{\star}$ University of California, Berkeley $\quad$
     	 $^{\dagger}$ Amazon Alexa AI}
\begin{document}
%
\maketitle
\input{sections/abstract}

\input{sections/introduction}
\input{sections/methods}
\input{sections/results}

\input{sections/conclusion}

\section{References}
\begingroup
  \def\section*#1{}
  \small
  \setlength{\bibsep}{4pt}
  \bibliographystyle{IEEEtranN}
  \bibliography{refs}
\endgroup

\end{document}

%% file: sections/abstract.tex
\begin{abstract}
Despite improvements to the generalization performance of automated speech recognition (ASR) models, specializing ASR models for downstream tasks remains a challenging task, primarily due to reduced data availability (necessitating increased data collection), and rapidly shifting data distributions (requiring more frequent model fine-tuning). In this work, we investigate the potential of leveraging external knowledge, particularly through off-policy key-value stores generated with text-to-speech methods, to allow for flexible post-training adaptation to new data distributions. In our approach, audio embeddings captured from text-to-speech, along with semantic text embeddings, are used to bias ASR via an approximate k-nearest-neighbor (KNN) based attentive fusion step. Our experiments on LibiriSpeech and in-house voice assistant/search datasets show that the proposed approach can reduce domain adaptation time by up to 1K GPU-hours while providing up to 3\% WER improvement  compared to a fine-tuning baseline, suggesting a promising approach for adapting production ASR systems in challenging zero and few-shot scenarios.
\end{abstract}
\begin{keywords}
speech recognition, transfer learning, fine-tuning, adaptation, context
\end{keywords}

\let\thefootnote\relax\footnotetext{For correspondence, direct questions to \url{davidchan@berkeley.edu}. Work done during an internship at Amazon Alexa AI. }

%% file: sections/introduction.tex
\section{Introduction}

One of the most challenging problems in automated speech recognition (ASR) is specializing large-scale models, particularly speech encoders, for downstream applications that often (a) have fewer labeled training examples, and (b) rapidly evolving distributions of speech data. The traditional approach to this problem is to frequently collect fresh data, which can be used to re-train and specialize models, leveraging tools such as domain-prompts \cite{Dingliwa2022}, incremental-learning \cite{Baby2022}, knowledge distillation \cite{Zhao2022}, hand-written grammars \cite{gandhe2018scalable}, or metric learning~\cite{papai2012,mahadevan2018} to reduce the impact of re-training the model for the downstream application. Unfortunately, for data that changes on a rapid basis, such as product listings or applications requiring per-customer specialization, such methods, while effective, are either inherently slow or remain computationally infeasible.

\begin{figure}
    \centering
    \includegraphics[width=\linewidth]{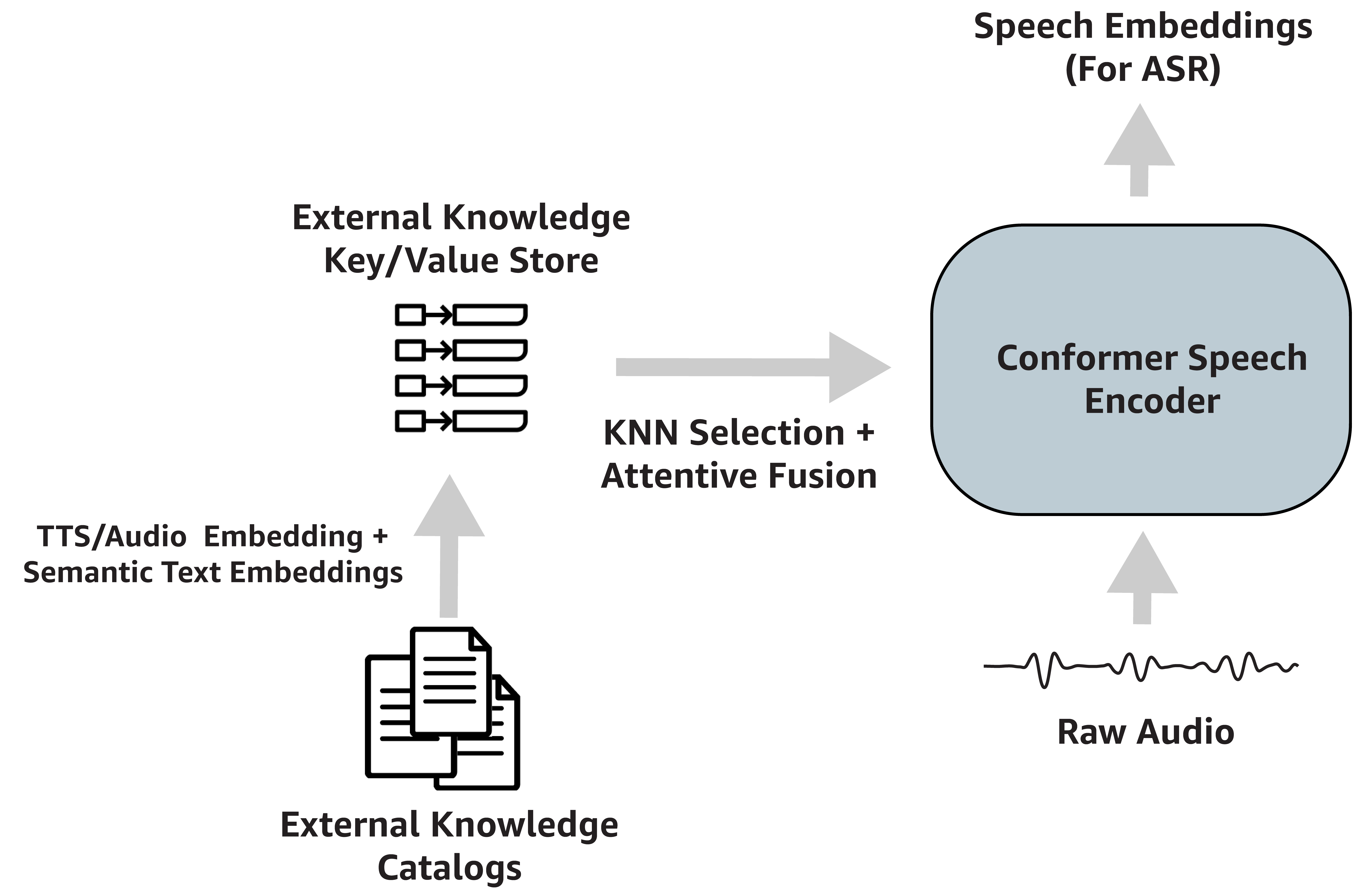}
    \caption{An overview of our method leveraging text-to-speech mappings for contextual ASR.
    Using data from a text catalog, we generate audio and text representations to generate mappings from audio key to text value. To leverage these mappings for ASR, we implement a K-Nearest Neighbors attention in the speech encoder during the fine-tuning (or training) phase.
    }
    \label{fig:teaser}
\end{figure}

In this work, we propose a method that leverages external text data catalogs -- large lists that can contain as much as 10 million specialized words or phrases -- to improve the performance of models during both the fine-tuning process, and when specializing an already fine-tuned model to a new dataset. Here are the key highlights of our approach: first, we generate a key-value external knowledge store that maps an audio representation of each text element of the catalog (usually consisting of 1M-10M examples) to a semantic representation of the text. Next, we train a model that leverages this external store by attending over retrieved key/value pairs, which we retrieve through approximate k-nearest neighbors. Relying on an external, constant, and off-policy key-value store means that this store can be updated during specialization, requiring only an updated list of phrases for each new model instead of additional fine-tuning.

Inspired by \citet{borgeaud2022improving} and \citet{wu2022memorizing}, we apply a context embedding approach with a focus on ASR, leveraging TTS-generated audio data and semantic text embeddings to bias the speech encoder of a conformer model. To the best of our knowledge, using TTS to encode textual context has not been explored in prior work.

Our key contributions are three-fold:
\begin{enumerate}
    \item We outline the first method (to our knowledge) to leverage large-scale text data for contextual biasing of the speech encoder.
    \item We show that our approach combined with an approximate K-NN lookup yields improved WER on ASR models, particularly in scenarios where encoded catalogs matches the target domain.
    \item We show that our approach can provide an accurate solution under the constraint of quick reactions to distribution changes (e.g., fast catalog updates for sporting events, changes in personal catalogs), without any model retraining.
\end{enumerate}

\section{Related Work}

Leveraging additional context to improve the performance of the \textit{decoder} and \textit{joint model}, particularly through deep fusion techniques in ASR transducers has been relatively well studied \cite{novotney2022cue, shenoy2021contextual, zhao2019shallow, liu2017dialog, jaech2018personalized, kim2018dialog,  lin2015hierarchical, williams2018contextual, munkhdalai2022fast}. The majority of these approaches are composed of two components: (1) a method for retrieving local contextual clues for an utterance and (2) a method for fusing these contextual clues with the joint model or decoder in the transducer stack, and differ based on how they implement these two components. Our work has several key differentiating factors. We primarily focus on deep-biasing of the speech encoder, rather than a shallow fusion of a context network with the speech decoder, or re-ranking of candidates produced by the language model. While in theory, approaches like \citet{sathyendra2022contextual} could be applied as deep fusion approaches, those directions are left unexplored in the referenced work, and indeed, as we demonstrate in \autoref{tab:layers}, deep fusion is an important differentiating factor in the quality of the contextual learning technique. Additionally, we focus on large contexts ($> 10K$ context entries), in an effort to explore applications in \textit{domain specialization}, whereas existing biasing methods are often focused on \textit{personalization}, and operate on contexts of at most $1K$ context entries (and in many cases, operate on $< 100$ context entries). Again, while in theory the above methods could be applied to larger contexts, we find that there is a strong implementational gap required, including work in scaling, and questions about efficiency (which we address in \autoref{sec:results}).

While late-stage fusion and biasing of the decoder and joint model have been well explored, biasing the speech encoder itself, particularly using early/deep-fusion approaches, remains an under-explored area of research. The closest work to our proposed model was presented by \citet{chen2019joint}, who use attention over a \textit{local} set of LSTM-based grapheme/phoneme embeddings to augment the audio encoder. They found that biasing the encoder with only 40 contextual text entities per utterance leads to improvements of up to 75\% on specialized test datasets. Similarly, \citet{sathyendra2022contextual} and \citet{chang2021context} demonstrate WER reductions when small ($<$100) contexts are fused in an attention-based process with both the speech and language model. Our method differs in that it is designed primarily for \textit{domain specialization}, whereas existing biasing methods are focused on \textit{personalization}. This is shown foremost in the scale of the catalogs -- while  in prior work, each utterance may have at most 100 utterances in their context, we leverage catalogs with up to 10M samples. Thus, our models are designed to compensate for \textit{general domain shift, rather than local personalized improvements to ASR performance}. Additionally, our work allows fully off-policy specialization. In existing works, context is re-encoded during each inference pass, leaving few opportunities for caching intermediate results. Our contexts are computed entirely offline; updates to the catalog do not impact model training, and thus, generating specialized catalogs can thus be done in a parallel work-stream from model development.

It is not unreasonable to believe that transformers in ASR will benefit from extended contexts. In the NLP field \citet{vaswani2017attention} showed that with longer memory attention contexts, transformers perform better, and a family of approaches, including \citet{dai2019transformer}, and \citet{child2019generating} have focused on increasing the length of the available context in each natural language sequence. The prevailing issue with long contexts is efficiency -- since transformers have quadratic scaling in the length of the context. To reduce the processing time, approaches such as \citet{dai2019transformer} and \citet{child2019generating} focused on computing gradients through only a subset of the full context, rather than the whole context to save memory and compute. Such an approach is codified by \citet{wu2022memorizing}, who recently demonstrated that expanding the context of standard text transformers through a large memory bank of cached external key-value pairs can lead to significant perplexity improvements on the standard language modeling task. \citet{wu2022memorizing} retrieve the most relevant context elements using a K-NN approach, and only back-propagate through these components. While the lookup may encourage some off-policy drift, the approach is effective, and allows for significantly increased performance, particularly in copy-tasks, which require the model to point at specific prior elements, which may not be accessible in models with smaller contexts.

Does extended context help when the context is external, or even, orthogonal to the current utterance? There seems to be some evidence to that effect, as outside of the standard ASR pipeline, it has been shown that models augmented with external memory generated from large-scale text data have the potential to outperform similarly sized models without external knowledge. \citet{borgeaud2022improving} recently demonstrated that leveraging external-knowledge lookup from a database of natural language sentences, can lead to efficiency improvements of up to 25x across a wide range of pure language tasks from language modeling to question-answering. Similarly, knowledge-augmented learning has been shown to be widely effective for QA \citep{marino2019ok, pan2019improving}, image captioning \citep{wu2017image} and other tasks \citep{tang2022improved, khandelwal2020nearest, goyal2022retrieval, xie2021survey, kumar2017use, tran2021deep}.

%% file: sections/methods.tex
\section{Methods}
\label{sec:methods}

\begin{figure}
    \centering
    \includegraphics[width=0.9\linewidth]{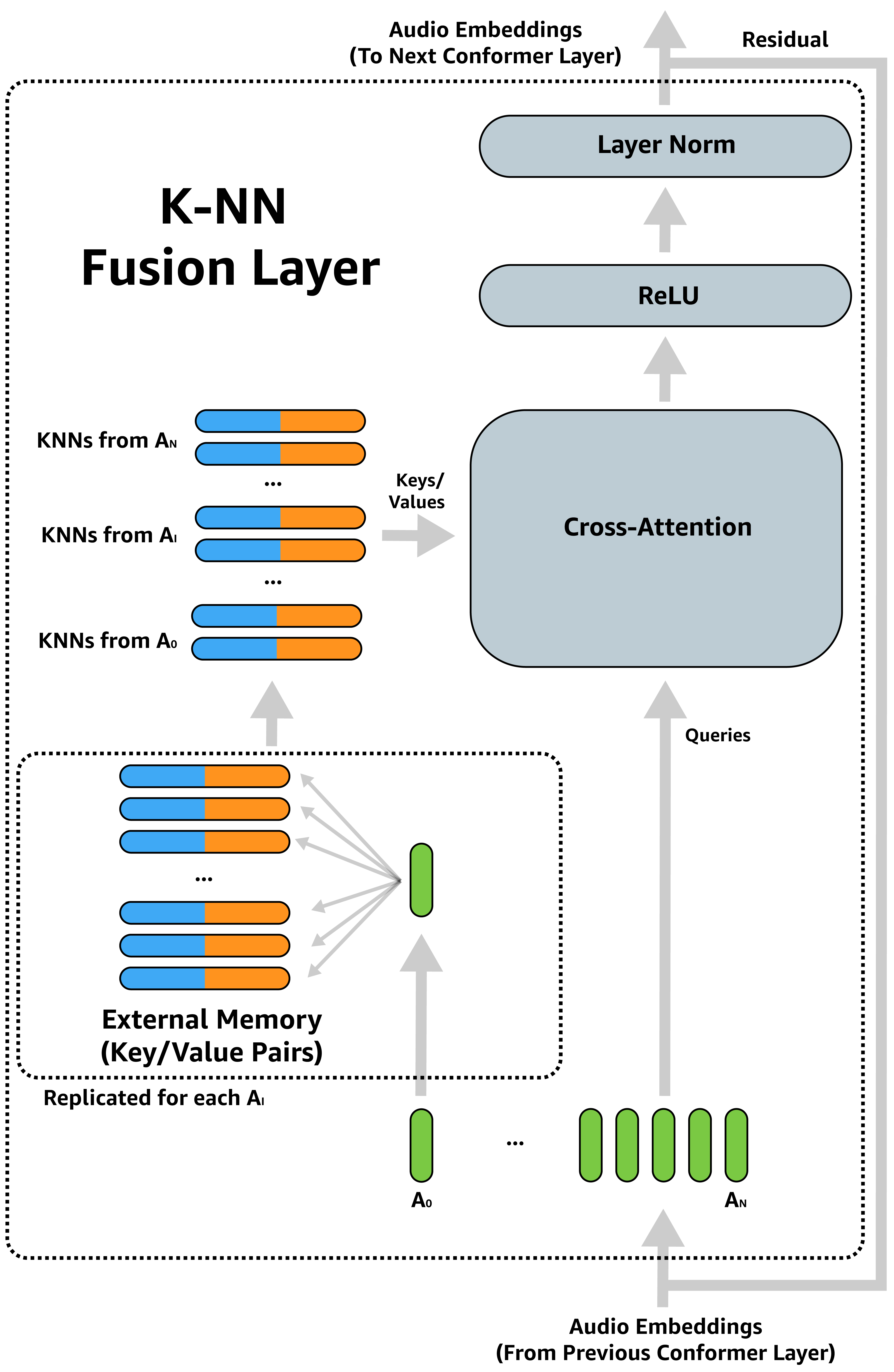}
    \caption{Overview of the K-NN fusion layer.
    For each audio frame embedding, we extract approximate KNNs using audio keys from our catalog. These KNNs form a context key/value store for a standard cross-attention layer \cite{vaswani2017attention}, where the queries are the incoming audio frame embeddings.
    }
    \label{fig:knn_fusion}
\end{figure}

\begin{figure}
    \centering
    \includegraphics[width=0.9\linewidth]{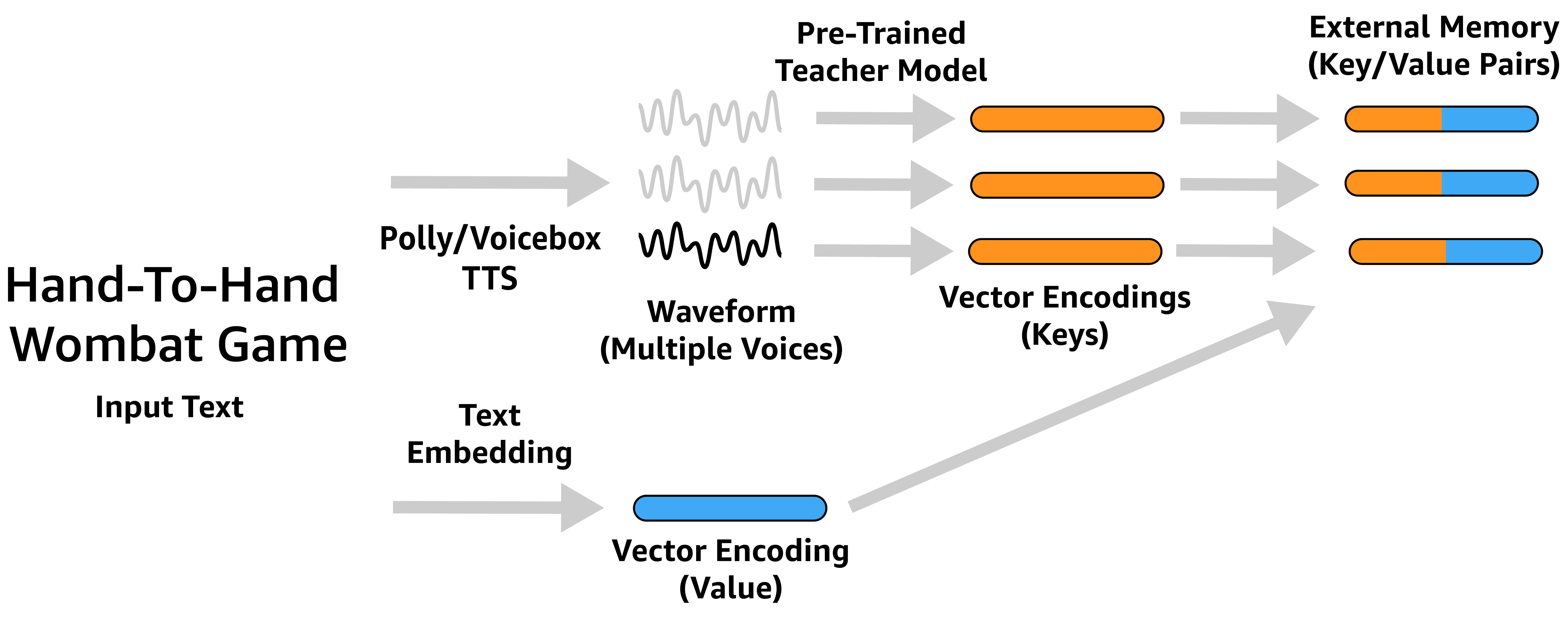}
    \caption{Overview of our text-catalog encoding process.
    For each catalog entry, we generate TTS-based audio encoding that forms the ``key" vector in the key-value pair. The value is a semantic text-embedding of the entry. Key/value pairs are assembled into the external memory, referenced in \autoref{fig:knn_fusion}
    }.
    \label{fig:embedding}
\end{figure}

An overview of our method is given in \autoref{fig:teaser}. Our approach consists of two key components: (1) A method for generating key-value mappings between the audible speech and a text representation of the catalog, which we call an ``external memory" and (2) An attention-based module for fusing the ``external memory" with the existing speech encoder. It is important that the external memory is able to be updated offline, and off-policy, as such a memory can be altered in a low-cost way, without incurring re-training costs.

\subsection{Generating the External Memory}

An overview of the external-memory generation process is shown in \autoref{fig:embedding}. Our approach generates the external memory consisting of audio-embedding key/text-embedding value pairs from a text-only catalog. To generate the audio-embedding key, we use text-to-speech (TTS) to generate waveform representations of the audio data, and then embed these waveform representations using the pre-trained speech encoder model. To generate the text-embedding values, we leverage off-the-shelf semantic text embedding methods, including 1-hot, GLoVE \cite{pennington2014glove} and BERT-style embedding \cite{devlin2018bert} approaches.

\subsubsection{TTS} In our work, we leverage two TTS modules to generate the audio for the audio-embeddings: the Amazon Polly TTS service\footnote{https://aws.amazon.com/polly/}, and an Alexa-AI Internal text to speech (TTS) library optimized for creating data for ASR model training and testing, which we will refer to as Multivoice-TTS as it can generate a number of voices. For both Amazon Polly and Multivoice-TTS we use ten voices, primarily drawn from the en-US and en-GB locales. 0.1 seconds of silence is inserted before and after each utterance.

\subsubsection{Audio Embedding} While audio embeddings for the external catalogue could be constructed in several ways, similar to \citet{wu2022memorizing}, we aim to make our audio-embeddings as close to on-policy self-attention embeddings as possible. Thus, we use the mean of the self-attention representations of our baseline model (no fine-tuning) at an intermediate layer, as our audio embeddings. 

\subsubsection{Text Embedding} In our work, we explore several methods of generate the text embeddings forming the value of the memory key-value pairs. For small catalogs, we explore learned one-hot embeddings, which are built during the training process. While such embeddings can lead to better performance (as they are built explicitly for each task), they are not scalable -- as they cannot be computed offline (and thus, cannot be inserted during test time). To generate scalable text embeddings, we explore two semantic text-emebdding approaches: GLoVE embeddings \cite{pennington2014glove}, which are built using word co-occurance probabilities, and BERT-style embeddings \cite{devlin2018bert}, which are learned from large statistical models. GLoVE embeddings are 300 dimensional, and computed using the publicly available vectors, and our BERT-style embeddings are computed using the \verb|all-MiniLM-L6-v2| model in the \verb|sentence-transformers| package \cite{thakur-2020-AugSBERT}.

\subsection{External Memory Fusion}

An overview of the external memory fusion process is given in \autoref{fig:knn_fusion}. The speech encoder in our proposed work is based on the Conformer encoder \cite{gulati2020conformer}, augmented with additional K-Nearest-Neighbor (KNN) fusion layers. In each KNN fusion layer, for each audio frame embedding $a_i$ of the utterance $A$, we query the external memory $E = (k_i, v_i), 1 \le i \le |E|$ for a set of $m$ nearest neighbors:
\begin{equation}
    \label{eq:nn}
    \mathcal{N}_{a_i} = \argmin_{N \subset E, |N|=m} \sum_{(k_j,v_j) \in N} || k_j - a_i ||_2^2
\end{equation}
We then construct the context for the layer as $\mathcal{C} = \cup_{a_i \in A} \mathcal{N}_{a_i}$. From $\mathcal{C}$ we can construct two matrices, $K_c \in \mathbb{R}^{m|A|,d_{\text{key}}}$ and $V_c \in \mathbb{R}^{m|A|,d_{\text{value}}}$, consisting of the keys and values respectively. The output of our K-NN fusion layer is then:
{
    \small
    \begin{equation}
        F(A, E) = A + \text{LN}\left(\text{ReLU}\left(\text{softmax}\left(\frac{(AW_q)K_c'}{\sqrt{d}}\right)(V_cW_v)\right)\right)
    \end{equation}
}
where LN is LayerNorm. Unfortunately, because we are working with large catalogues, the computation of \autoref{eq:nn} can be very expensive. Thus, instead of computing exact nearest neighbors, we rely on approximate nearest neighbors, which can be computed much more efficiently. To efficiently extract approximate nearest neighbors from our large-scale catalogs, we leverage the FAISS \cite{johnson2019billion} library to generate Optimized Product-Quantization-transformed keys (64 dimension) \cite{ge2013optimized}, which are searched using an Hierarchial Naviagable Small Worlds (HNSW) index with 2048 centroids encoded with product-quantized fast-scan \cite{malkov2018efficient}. Such an approach leads to only a 15\% increase in forward-pass latency, even when running with catalogs with over 7M key/value pairs.

\subsection{Experimental Design}

\subsubsection{ASR Base Model} Although in practice our method could be applied to many different speech encoders, we use the Conformer encoder \cite{gulati2020conformer}. For the decoder, we use a 1-layer LSTM decoder with 320 hidden dimension, with no explicit pre-trained language model. While we explore several encoder sizes, we primarily follow \citet{gulati2020conformer} for Librispeech and use a 16 layer encoder with a hidden dimension of 144 (10.3M Params). For internal Alexa-AI data, we use a conformer model with 208.37M parameters.  All models leverage ReLU activations, batch-normalization, and dropout of $0.1$. For the ASR tokenization, we use a sentence-piece model \cite{kudo2018sentencepiece} with a vocab size of 640 (librispeech) and 4096 (internal).

\subsubsection{Catalog Data Sources} In our work we explore several different catalog data sources. For Librispeech, we build a simulated catalog using the 2500 rarest tokens present in either the training or test datasets. Building a unique catalog for both the training and the test data allows us to explore how well the model performs under distribution shift of the catalog at test time. Our internal Alexa catalog focuses on assistant queries in a media domain, and consists of 15K movie titles.

\subsubsection{Training Details} For Librispeech, the model is implemented in Tensorflow, and is trained using 24 Nvidia V-100GPUs for 120 epochs with a batch size of 2048 and the Adam optimizer, with a learning rate of $3e^{-4}$. For the Alexa-AI datasets, the model is fine-tuned using 104 Nvidia V-100 GPUs for $30$ epochs with a batch size of $832$ and the Adam optimizer, with a warmup/hold learning rate schedule with $10,000$ warmup steps and a maximum learning rate of $5e^{-3}$.


%% file: sections/results.tex
\section{Results \& Discussion}
\label{sec:results}

In this work, we present results on two datasets - Librispeech \cite{panayotov2015librispeech}, a dataset consisting of 960 hours of relatively clean, annotated ASR data, and an internal Alexa dataset focused on media-centric queries.

\begin{table}
\footnotesize
\caption{\small Word Error Rate on  Librispeech data with a small (10.3M param) model for several choices of TTS, Text Embeddings, and NNs/Frame (K). MV-TTS refers to Multivoice-TTS.}\label{tab:ls}
\begin{tabularx}{\linewidth}{lrrrrr}
	\toprule
		\textbf{Catalog}& \textbf{TTS} & \textbf{Text} & \textbf{K} & \textbf{test-clean} & \textbf{test-other} \\
		
	\midrule
	\textbf{Baseline} &&&& 5.77  & 13.34 \\
	\midrule
	\textbf{Train} & Polly    & 1-Hot  & 4  & 5.75 \tiny{(0.34\%)} &  13.30 \tiny{(0.29\%)}  \\
	    & Polly    & 1-Hot  & 8  & 5.72 \tiny{(0.86\%)} &  13.19 \tiny{(1.10\%)}  \\
	    & Polly    & 1-Hot  & 16 & 5.71 \tiny{(1.03\%)} &  13.15 \tiny{(1.42\%)}  \\
	    & Polly    & BERT   & 8  & 5.74 \tiny{(0.52\%)} &  13.26 \tiny{(0.60\%)}  \\
	    & MV-TTS & 1-Hot  & 8  & 5.52 \tiny{(4.33\%)} &  12.96 \tiny{(2.84\%)}  \\
	    & MV-TTS & BERT   & 8  & 5.68 \tiny{(1.63\%)} &  13.05 \tiny{(2.18\%)}  \\
	\midrule
	\textbf{Test} & Polly    & GLoVE  & 8  & 6.33 \tiny{(-8.84\%)} &  14.56 \tiny{(-9.15\%)}  \\
	    & Polly    & BERT   & 8  & 5.71 \tiny{(1.03\%)} &  13.24 \tiny{(0.75\%)}  \\
	    & MV-TTS & GLoVE  & 8  & 6.15 \tiny{(-6.17\%)} &  14.32 \tiny{(-6.84\%)}  \\
	    & MV-TTS & BERT   & 8  & \textbf{5.34} \tiny{(8.05\%)} &  \textbf{12.84} \tiny{(3.86\%)}  \\
	\bottomrule
\end{tabularx}
\end{table}

\begin{figure}
\begin{tikzpicture}
\begin{axis}
[
    width=\linewidth,
    height=0.4\linewidth,
    xlabel=\% of Test-Clean bigrams overlapping catalog,
    ylabel=WER,
    ytick distance=0.2,
    tick label style={font=\footnotesize}
]
    \addplot
    plot[orange, mark options={orange}] coordinates {
        (0, 5.76)
        (10, 5.75)
        (20, 5.76)
        (30, 5.71)
        (40, 5.63)
        (50, 5.24)
        (60, 5.01)
        (70, 4.95)
        (80, 4.86)
        (90, 4.72)
        (100, 4.73)
    };
\end{axis}        
\end{tikzpicture}
\caption{Librispeech test-clean WER over differing test catalogs. As the percentage of bigrams in the test catalog overlapping with the test dataset increases, the performance of the catalog-augmented model increases as well.}
\label{fig:cats}
\end{figure}
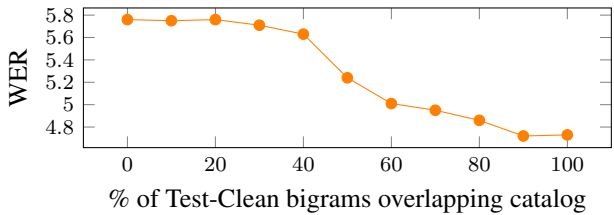

\begin{table}
\small
\caption{\small Relative Librispeech test-set WER improvement for models augmented with catalog data in different layers Model uses Multivoice-TTS, BERT Embeddings and 8 NNs/Frame.}\label{tab:layers}
\begin{tabularx}{\linewidth}{Xrrrrrr}
	\toprule
		\textbf{Dataset} & \textbf{1} & \textbf{3} & \textbf{12} & \textbf{16} & \textbf{3,12} & \textbf{all} \\
	\midrule
	clean & 1.02\% & 3.65\% & 6.65\% & 2.63\% & 7.79\% & 8.05\% \\
	other & 0.71\% & 2.88\% & 2.97\% & 1.08\% & 3.41\% & 3.86\% \\
	\bottomrule
\end{tabularx}
\end{table}

\begin{table}
\small
\caption{\small Relative Librispeech test-set WER improvement over baseline fine-tuning using differing model parameters with Multivoice-TTS, BERT, and 8 NNs/Frame.}\label{tab:params}
\begin{tabularx}{\linewidth}{Xrrrrr}
	\toprule
		\textbf{Dataset} & \textbf{5M} & \textbf{10M} & \textbf{50M} & \textbf{100M} & \textbf{300M} \\
	\midrule
	clean & 28.9\% & 8.05\% & 4.28\% & 1.66\% & 0.08\%  \\
	other & 19.3\% & 3.86\% & 2.65\% & -0.07\% & 0.01\%  \\
	\bottomrule
\end{tabularx}
\end{table}

\begin{table}
\small
\caption{\small Alexa-AI Performance. T-C: Time for Catalog Generation. T-FT: Time for fine-tuning.  Multivoice-TTS, BERT, and 8 NNs/Frame.}\label{tab:ftv1}
\begin{tabularx}{\linewidth}{Xrrr}
	\toprule
		\textbf{Model/Test Data} & \textbf{T-C} {\tiny (min)} & \textbf{T-FT} {\tiny (GPU-Hours)} & {\tiny rel.} \textbf{TTS WER}  \\
	\midrule
	$B_{\text{FT}}$/Train-TTS  & 0 & 2048 & 7.1\% \\
	$B_{\text{cat}}$/Train-TTS  & 33 & 1600 & 6.8\% \\
	$B_{\text{FT}}$/Test-TTS  & - & - & 0.52\% \\
	$B_{\text{cat}}$/Test-TTS  & - & - & 4.12\% \\
	$B_{\text{FT} + T}$/Test-TTS  & 0 & 1024 & 19.66\% \\
	$B_{\text{cat} + T}$/Test-TTS  & 28 & 0 & 21.27\% \\
	\bottomrule
\end{tabularx}
\end{table}

\subsection{Librispeech} Our key results are shown for Librispeech in \autoref{tab:ls}. We can see that overall, augmenting models with additional data leads to stronger performance than models without external data. For Librispeech, when training with the train catalog and testing with the test catalog, we get strong transfer performance, exceeding that of when we use the training catalog for both training and testing, suggesting additional zero-shot specialization. While 1-hot vectors outperform BERT vectors, we must train these vectors for each catalog, leading to an inability to do test-time specialization. BERT outperforms GLoVE in all cases (with GLoVE causing regressions on test-time specialization). \autoref{fig:cats} demonstrates that our method can capture and apply domain data from the catalogs. In this experiment, the model is trained with a catalog containing 300K training-set unique bigrams, and we show the performance of this model using ten test catalogs, each consisting of 30K bigrams, taken either from the test set or dev set. As the fraction of bigrams in the test data that are available in the test catalog increases, the performance of the model improves -- showing our approach can use the information in test catalogs effectively in a zero-shot learning setup. \\

\paragraph{Ablations:} \autoref{tab:layers} explores the performance of our model when placing the external knowledge augmentation at different layers of the model. While making external knowledge available to all layers is the most effective approach, we find that such an approach is latency-prohibitive, as it increases the latency of a forward pass of the model by about 85\%. Using a single layer increases latency by only about 15\%, while two layers increase latency by about 23\%. \autoref{tab:params} explores the performance of the method on Librispeech as we increase the number of parameters in the model. As we increase the number of parameters, the gains provided by external memory decrease.

\subsection{Alexa} To further validate our method, we additionally explore a real-world simulation of our model's ability to generalize to test data. We started with a baseline model $B$ (See: \autoref{sec:methods}), and trained two derived models: $B_{\text{FT}}$, fine-tuned on both the TTS Catalog for Alexa ($\mathcal{C}$, \autoref{sec:methods}) and an additional 120K hours of de-identified Alexa data, $\mathcal{D}$, and $B_{\text{cat}}$, which applies our method fine-tuned on $\mathcal{D}$, with catalog $\mathcal{C}$. The results (\autoref{tab:ftv1}, rows 1/2) demonstrate that even with significantly fewer GPU hours, our approach achieves similar WER. When we transfer to the test dataset (without updating the catalog), we see in \autoref{tab:ftv1} (rows 3/4) that our trained model achieves better performance, suggesting that the model has learned to generalize better than the model trained with fine-tuning alone. 

Finally, we update our fine-tuned and catalog models to include the test data. The test data is incorporated into the fine-tuned model through additional GPU-based training, while the test data is incorporated into the catalog model through catalog generation and concatenation. \autoref{tab:ftv1} (rows 5/6) further demonstrates that even with \textit{no additional GPU training} our approach ($B_{\text{cat}+T}$) can achieve similar performance to the fine-tuning ($B_{\text{FT} + T}$) approach.

%% file: sections/conclusion.tex
\section{Conclusion}

This paper introduces the first approach for large-scale contextualization of speech-encoder representations using text-only catalog data. While this paper is a good first step towards contextualized speech encoders, problems like investigating embeddings for the catalogs, leveraging grapheme/phoneme embeddings, etc. remain interesting directions of future work. This approach provides a natural way to combine external memory for addressing distribution shifts when having OOV words in dev/test,  ensuring recognition of rare words in training data, handling personalization, and using pronunciation instead of TTS -- we would like to evaluate these features of the approach on real-world data. 

